\documentclass[3p,times,twocolumn,authoryear]{elsarticle}
\usepackage{lineno,hyperref}
\usepackage{amsmath}
{}
{}
{}
\usepackage{float}
\usepackage{scrextend}
\usepackage{caption}
\usepackage{multicol,lipsum}

\journal{COMPUTERS $\&$ SECURITY}

\begin{document}

\begin{frontmatter}

\title{ProAPT: Projection of APT Threats with Deep Reinforcement Learning}

\author{Motahareh Dehghan $^1$} 
\author{Babak Sadeghiyan $^2$}
\author{Erfan Khosravian $^3$}
\author{Alireza Sedighi Moghaddam $^2$}
\author{Farshid Nooshi $^2$}

\address{$^1$ Tarbiat Modares University, Tehran, Iran \\}

\address{$^2$ Amirkabir University of Technology, Tehran, Iran \\}

\address{$^3$  Payame Noor University, Tehran, Iran}

\begin{abstract}

The highest level in Endsley's situation awareness model is called projection, when the status of elements in the environment in the near future is predicted. In cybersecurity situation awareness, the projection for an Advanced Persistent Threat (APT) requires to predict the next step of the APT. \\
The threats are constantly changing and becoming more complex. As supervised and unsupervised learning methods require APT datasets for projecting the next step of APTs, they are unable to identify unknown APT threats.\\
In reinforcement learning methods, the agent interacts with the environment, and so it might project the next step of known and unknown APTs. So far, reinforcement learning has not been used to project the next step of APTs.\\
In reinforcement learning, the agent uses the previous states and actions to approximate the best action of the current state. When the number of states and actions is abundant, the agent employs neural network which is called deep learning to approximate the best action of each state.\\
In this paper, we present a deep reinforcement learning system to project the next step of APTs. As there exists some relation between attack steps, we employ Long- Short Term Memory (LSTM) method to approximate the best action of each state. In our proposed system, based on the current situation, we project the next steps of APT threats.\\
We have evaluated our proposed deep reinforcement learning system on the DAPT2020 dataset. Based on the evaluations performed on the mentioned dataset, six criteria F1, Accuracy, Precision, Recall, Loss, and average time were obtained, which are 0.9533, 0.9736, 0.9352, 0.97, 0.0143, and 0.05749(seconds) respectively. Although there is no previous research on using reinforcement learning for APT projection, our results compared to the previous supervised and unsupervised methods proposed for multi-step attack projections indicate appropriate functioning. \\
  
\end{abstract}

\begin{keyword}
Situation Awareness \sep Advanced Persistent Threats \sep Projection \sep Deep Reinforcement Learning \sep LSTM \sep DAPT2020
\end{keyword}
\end{frontmatter}


\section{Introduction}
Expansion of capacities, opportunities, threats and applications in cyberspace have increased the motivation and efforts to better detection and perception. Information sources that are stored on computers are worthy targets for terrorist and criminal groups, hostile governments, spy agencies, political and economic rivals. Although the application of information and communication technology has made it possible to communicate and manage efficiently, this technology has enabled hostile individuals and organizations to identify and exploit vulnerabilities in seemingly secure computer networks. Protection and maintenance of cyberspace is a more complex process than traditional information and communication networks.  \\
In cyberspace, threats have a complex form and include internal and external attackers with different skill levels. Currently, attackers use automated tools to exploit and control target systems remotely. When systems are infiltrated, attackers use them to carry out more attacks and achieve their next targets. In this case, cybersecurity situation awareness is important for the analysis of cyberspace, and detection of ever-changing threats. \\
Situation awareness is a cognitive process that can percept and comprehend the current situation, and project the near future. Then, based on the obtained awareness, plan, decision, and act can be performed.
There are different definitions for situation awareness. One of the most famous of which provided by Mica Endsley in 1995 [1]: " Situational awareness or situation awareness (SA) is the perception of environmental elements and events with respect to time or space, the comprehension of their meaning, and the projection of their future status." \\
This definition makes a subtle distinction between three levels of situation awareness, i.e. perception (including observation), comprehension, and projection (including prediction). Its lowest level is observation and perception, and the highest level is projection of the near future, i.e. the projection of the current situation into the future in an attempt to predict the evolution of the tactical situation. \\
The highest level in Endsley's situation awareness model is called projection, when the status of elements in the environment in the near future is predicted. In cybersecurity situation awareness, the projection for an Advanced Persistent Threat (APT) is to predict the next step of the APT. \\
Especially in the area of cybersecurity awareness, the prediction of future cyberattacks or the next step of the current attack is very important. \\
Attacks usually take long periods for cyber espionage or sabotage, and involve a lot of reconnaissance, exploitation, and obfuscation activities. In general, the future activity of the attack is predicted through current destructive activities. Such projections require analyzing potential attack paths based on the network and system vulnerabilities and knowledge about attacker behavioral patterns. \\
Projection of Advanced Persistent Threats (APT) requires modeling the process of penetration over the time. This model is much broader than traditional definitions in the field of intrusion detection, in which the focus is on studying the vulnerabilities and exploitation of the system. Reinforcement learning can be used for the projection of APTs. In reinforcement learning problems, some agents interact with the environment through trial and error and learn to select the optimal action. In this type of learning, there are no external observers, and the agents interact with the environment alone, learn and gain experience and receive a reward [2]. \\
In reinforcement learning, agents are equipped with sensors that can get features of the environment. These features describe the environment states. Then, the agents influence the environment by performing some actions. Therefore, the agents receive rewards in the next time based on the chosen action. When the number of states and actions are abundant, the agent employs neural networks to approximate the best action of each state. To know more about the details of reinforcement learning, an interested reader is referred to [2].\\
As we have stated, reinforcement learning can be used to project the next step of APTs. In reinforcement learning, the agent interact with the environment for APT projection. Hence, it is suitable for unknown APT projection. Moreover, deep learning can be used to approximate the best action of each state.\\
In this paper, we propose a deep reinforcement learning system for APT projection. We determine the important parameters of our proposed system. Due to the great impact of these parameters on the result of projection, we train the agent with different values of parameters and evaluate the results to assign the best values to them. Finally, we evaluate our proposed system and calculate the metrics Accuracy, F1, Precision, Recall, Loss, and average time. \\
In the next section, we review previous researches in the field of attack prediction and projection. Then, we present our proposed system for APT projection with deep reinforcement learning. Afterward, we evaluate our proposed system and determine the best values for its parameters. 
\section{ Related Work }
Projecting multi-step attacks requires modeling the process of penetration over the time. This model is much broader than traditional definitions of intrusion detection, in which the focus is on studying the vulnerabilities and exploitation of the system. In the late 1990s, Cohen introduced the first framework for modeling network attacks. He used a cause-and-effect model to deduce the existence of 37 characteristics of attacks, 94 cyber and physical attacks, and 140 defense mechanisms. He also published his simulation results in the form of reports [3, 4].  Cohen's work, along with several other research work in the late 1990s and early 2000s, has provided a comprehensive understanding of the types of cyberattacks and their impact on network systems [5-7]. \\
Initial work on attack modeling, or more precisely in the field of attack classification, led to research on the alert correlation or attack plan recognition [8-17]. \\
In the alert correlation, the general idea is that the hypothetical attack model, called the attack graph or attack plan, is first constructed, and based on the analysis of these models, the attack prediction or projection is performed [18-33]. \\
An alternative approach instead of using an attack graph is to use a Dynamic Bayesian Network [34-38]. The distinction between the attack graph and the dynamic Bayesian network is that the attack graph is rule- based, while the Dynamic Bayesian Network approach is probabilistic and data-driven. The rule- based feature of the attack graph empowers modeling and analyzing vulnerabilities and exploit codes in a more specific and detailed manner.\\
To identify learning requirements and predict ongoing network attacks in an online method, Fava et al. [39] proposed a projection system based on the Markov model with variable length. For the first time, Fava et al. used the variable length Markov model to project attacks. The variable-length Markov model is well compatible with online learning systems because it has excellent computational efficiency compared to the hidden Markov model and high flexibility compared to the fixed-length Markov model [40-54].  \\
Game theoretical approaches are similar to graph and Markov-based approaches. In Game Theoretical approaches, play is a model of interaction between attacker and defender. Unlike previous approaches, the goal of this approach is to find the best strategy among players [55-57].\\
Machine learning and data mining approaches such as Neural Networks, Support Vector Machines, Association Rule Mining and, etc. are also provided for intrusion detection and attack projection [58-79]. \\ 
Other methods have been proposed to predict and project attacks using continuous models such as Time Series and Grey Models [80-94]. These models are very suitable for predicting network security situations. \\
As stated above, reinforcement learning has not been used to predict and project the next steps of APT threats. Reinforcement learning helps to project unknown APTs by interacting with the environment. 
In reinforcement learning, when the number of states and actions is abundant, the agent employs neural networks to approximate the best action in each state. Hence, we present a deep reinforcement learning system for APT projection. In our proposed system, based on the current situation, we project the next steps of APT threats. This contribution helps to select appropriate and timely countermeasures against APTs.
\section{Preliminary Definitions}
\subsection{Advanced Persistent Threat (APT)}
APT is an increasingly sophisticated multi-step attack by hostile organizations to gain access to defense, financial, and other targeted information from governments, corporations, and individuals. Maintaining a foothold in these environments to enable future use and control, and modifying data to disrupt performance in their targets are the goal of APTs[95].\\
Given the description of APTs and their extent, in this paper, we propose a deep reinforcement learning system to project the next steps of APTs. Hence, in the next section, we describe deep reinforcement learning. 
\subsection{Deep Reinforcement Learning}
Reinforcement Learning (RL) is one of the areas in machine learning that aims to improve the behavior of intelligent agent based on the signals it receives from the environment [2]. \\
In RL, an agent interacts with the environment, learns through trial and error, and chooses the optimal action to achieve the goal. There are no external observers in this type of learning, and the agent interacts with the environment alone, learns and gains experience, and receives a reward. To know more about the details of reinforcement learning, an interested reader is referred to [2].\\
Any RL problem is defined by three important components, i.e. states, actions and rewards. States are a representation of the current world or environment of the task. Actions are something a RL agent can perform to change these states. Moreover, rewards are the utility the agent receives for performing the right actions. So the states tell the agent what situation it is in currently, and the rewards signal the states that it should be aspiring towards. The aim, then, is to learn a policy, something that tells you which action to take from each state to try and maximize reward [2].
In RL, Q-learning is the most effective method and is widely used in the literature [2, 96 , 97]. We state the details of Q- learning in the next section. 
\begin{figure*}
\begin{center}
\includegraphics[scale=0.7]{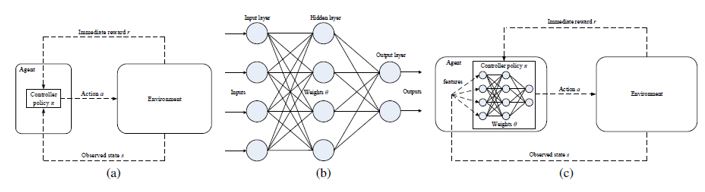}
\small \caption{(a) Reinforcement learning (e.g. Q-learning), (b) Artificial neural network, and (c) Deep Reinforcement Learning (e.g. Deep Q-learning) [96]}
\label{Figure 1} 
\end{center}
\end{figure*}
\nopagebreak
\subsection{Q- Learning}
As stated, one of the most common algorithms in RL is the Q-learning algorithm. In Q- learning, the value function is the sum of the rewards received by the agent and the selection of the best action based on the policy. The agent updates the value functions using relationship (1) until it converges to the optimal value. It should be noted that this algorithm does not need to have a model of the environment. In addition, in each episode, when $s_{t+1}$ is a final state, the Q function value for it will never be updated and maintain its initial value [2, 96, 97].\\
$Q(s_t,a_t ) \leftarrow Q(s_t , a_t )    + \alpha [ r_{t+1}+ \gamma _{max_a}  Q(s_{t+1},a_{t+1} )  - Q(s_t,a_t )  ]      \  \  \  \   \                              (1)$
The following pseudo-code shows Q-learning algorithm [96]:
\begin{enumerate}
\item Initialize Q(s,a) arbitrarily
\item Repeat (for each episode)
\item \ \ \ \ Initialize \textit{s}
\item \ \ \ \ Repeat (for each step of episode)
\item \ \ \ \ \ \ \  Choose \textit{a} from \textit{s} using policy derived from \textit{Q}(e.g., $\epsilon-greedy$)
\item \ \ \ \ \ \ \ \ Take action \textit{a}, observe \textit{r,s'}
\item $Q(s_t , a_t) \leftarrow Q(s_t , a_t) + \alpha [r_{t+1} + \gamma max_a . Q(s_{t+1},a_{t+1})-Q(s_t,a_t)]$ 
\item \ \ \ \ \ \ \ \ $s \leftarrow s'$
\item \ \ \ \ Until \textit{s} is terminal; 
\end{enumerate}
In Q- learning algorithm, if the space of states and actions is limited, the Q- table is used. In the Q-table, for each state 's' and action 'a', the value is specified. However, if the space of states and actions is wide and consumes a lot of space, function approximation methods are used. The use of neural networks is one of the most common and practical methods for approximating the value function, which is called deep learning. By the combination of deep learning and Q- learning, we have deep Q- learning, as demonstrated in Fig. 1. To know more about the details of Q- learning, an interested reader is referred to [96, 97].\\
When the number of states and actions is high, it is necessary to use the approximation function for estimating the value function.\\
In sections 3.4 and 3.5, we describe two types of neural networks that we employ as approximation function.
\subsection{Recurrent Neural Networks}
As stated above, neural networks can be used to approximate the value functions in reinforcement learning. Neural networks are a class of models with multilayered structures. Convolutional neural networks (CNN) and recurrent neural networks (RNN) are common types of neural networks [96, 97]. RNN is a form of neural network that has internal memory. RNN is inherently recursive because it executes the same function for all input data, but the output of the current (input) data depends on the previous calculations. After production, the output is copied and sent back to the recurrent network. Recurrent neural networks use their internal state (memory) to process sequences of inputs, as  illustrated in Fig. 2. In other neural networks, inputs are independent of each other, but in recurrent neural networks, the inputs are interrelated [96, 97].
\begin{figure}[H]
\begin{center}
\includegraphics[scale=0.7]{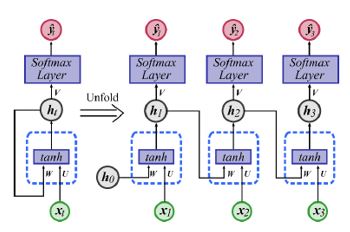}
\small \caption{Recurrent Neural Networks [97]}
\label{Figure 2} 
\end{center}
\end{figure}
RNN is essentially used when the context is important, i.e. when previous iterations or instances can affect the current iteration. As a common example of these types of fields, text can be mentioned. In the text, a word can be analyzed only in the context of the previous word or sentence. \\
The disadvantages of RNN include gradient descent and its problems, difficult training, and failure to process long sequences of inputs in case of using tanh or relu activation functions. Problems in RNN can be solved by Long-Short Term Memory (LSTM) method [96, 97].
\subsection{Long- Short Term Memory}
LSTM is a modified version of recurrent neural networks that facilitate the recall of past data. The gradient descent problem in the RNN has also been solved in these networks. These networks train the model using backward propagation, as demonstrated in Fig. 3. 
\begin{figure}
\begin{center}
\includegraphics[scale=0.7]{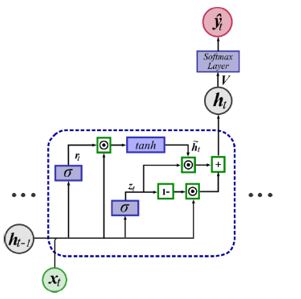}
\small \caption{Long Short Term Memory [97]}
\label{Figure 3} 
\end{center}
\end{figure}
\section{Proposed Deep Reinforcement Learning  for APT Projection}
APT threats have different steps to achieve the ultimate goals. The detailed description of different APT phases are as follows [98]:
\begin{itemize}
\item Reconnaissance: scan applications and networks
\item Establish Foothold: download or install malware, R2L, command and control communication
\item Lateral Movement: credential compromise, privilege escalation (U2R)
\item Internal Reconnaissance: Same as Reconnaissance above, just from a different source in search of data.
\item Data Exfiltration: Uploading to Google Drive, Dropbox, AWS, or any such cloud. 
\item Cover Up: Deletion of log files, modification of log files, etc.
\end{itemize}
In this research, we propose a deep reinforcement learning system (DRLS) based on Q- learning and LSTM to project the next step of APT threats. As there exist some relations between the attack steps, we use LSTM for value function approximation. As mentioned before, LSTM is a modified version of RNN and facilitates the recall of past data and solves the problems of RNN. We use LSTM to keep the previous states over long periods.\\
The APT projection problem can be considered as a Markov Decision Process. Because, detection of normal or abnormal behavior at the current time step will alter the environment. Changing environment will also influence the next decision. Hence, it is natural to adapt this problem to the framework of RL [2]. We describe the DRLS for the APT projection problem as follows:\\
We demonstrate each state by some features such as the source IP address, destination IP address, source port number, destination port number, timestamp, attack type, etc.  The agent receives the current state and selects the best action based on the $\epsilon$ -greedy  policy. Indeed, the agent receives the correlated alerts and selects the next step of attack. The reward is a 1/0 reward for a correct/ incorrect attack prediction. We use Q-learning algorithm to learn the agent. To approximate the Q function, we employ LSTM, as there exist some relations between attack steps. A Q function provides the maximum expected reward at a specific state and action. \\
To reduce the time of learning, testing, and evaluation, we employ APT datasets instead of interacting with the environment. Although employing datasets increases the speed of learning and testing, interacting with the environment is suitable for predicting unknown APTs. In our proposed model the agent can employ datasets or interact with the environment to project the next step of APTs.\\
Based on Fig. 4, we divide the input dataset into sections and select the index randomly. Then, from the selected index, we consider the N number of data as training data. Each Training data, as input for LSTM, includes the features of the alerts. In addition, the second part of the training data includes the label of the next step of the attack. For example, $S_0$ represents the attack step in the time of $t_0$, $a_1^*$ expresses the attack label at time $t_1$ and for the state $S_1$. Since we want to recognize the next step of the attack in the DRLS, we consider the next step label in each state and use it to determine the reward for it.
\begin{figure*}
\begin{center}
\includegraphics[scale=0.7]{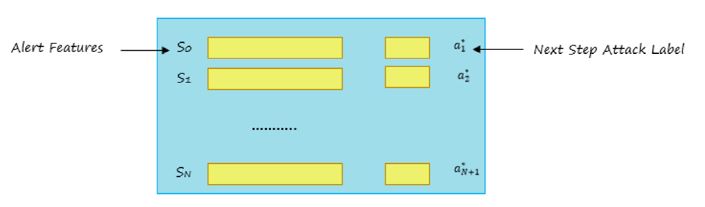}
\small \caption{Data preparation for proposed DRL}
\label{Figure 4} 
\end{center}
\end{figure*}
\begin{figure*}
\begin{center}
\includegraphics[scale=0.15]{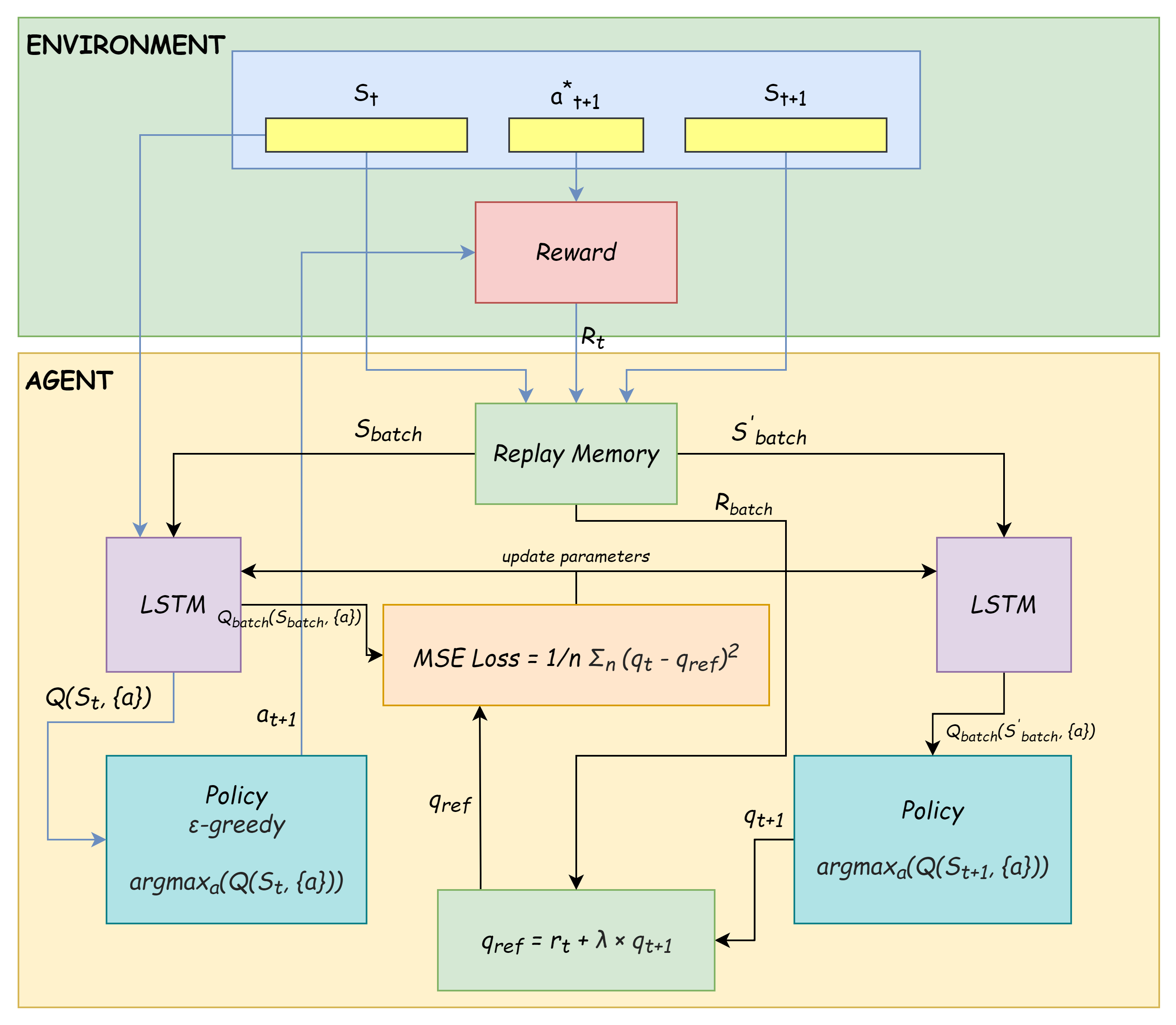}
\small \caption{The Architecture of the Proposed DRLS (Deep Reinforcement Learning System)}
\label{Figure 5} 
\end{center}
\end{figure*}
\nopagebreak
In Fig. 5, a DRLS is demonstrated to predict the next step of the attack. As mentioned before, we give data from an APT dataset as input to DRLS.\\
Based on the input data, the agent learns how to predict the next step of attacks. Input data consists of three parts. The first part expresses the state at time t. This part includes the features of the correlated alerts at time t. The second part is the data label in step t+1. This part shows the attack label in step t+1. The third part describes the state at time t+1. That is, features of correlated alerts at time t+1. \\
The first part of the input is entered into LSTM neural network to approximate the value function of different actions for the state at time t. In this context, LSTM approximates the value function for the next step of the ongoing attack. We display the approximated value with $\hat{a_{t+1}}$, which has the value $\hat{q_t}$. Then, based on the $ \epsilon$ -greedy policy, the action with the highest value function is selected by a probability of $\epsilon$. Finally, the approximated value $(\hat{a_{t+1}})$ is compared with the main label of the next step of the attack, which is the second part of the input data $(a^*_{t+1})$. If the comparison result is equal, the reward +1 is given to the agent; otherwise, the reward 0 is given.\\
The third part of the input is used to calculate the error function and update the LSTM. So that the state-expressing features at time t+1 are entered in the second LSTM (right part in Fig. 5) for approximation of value functions for different actions. \\
At this point, the policy is the selection of action with the greatest value. In our problem, actions are the next step of attacks. Then, the obtained value is used to calculate the Mean Square Error Loss ($Loss \ = \ \frac{1}{n} \ \Sigma(\hat{q}_t - q_{ref})^2$ ) between the Q-value approximated by the LSTM for the state at time t and a reference value ($q_{ref}$). The reference value ($q_{ref} \ = \ r_t \ + \ \lambda \ \times \ \hat{q}_{t+1}$) is obtained by adding the reward at time t ($r_t$) to the Q-value for the state at time t+1 multiplied by a discount factor ($\lambda$).
\section{Experiments}
In this section, we will determine the efficiency of our proposed DRLS by performing experiments on a dataset of advanced persistent threats, called DAPT2020 [98]. \\
DAPT2020 was generated over five days, which can be considered equivalent to collecting network traffic over three months in a real-world scenario. This dataset was created to give researchers the ability to understand anomalies, and relationships between different attacks and find any hidden correlations to detect APTs in their early steps. \\
For performing our experiment, the dataset was first loaded and integrated. Then, it was cleaned. However, DAPT2020 does not have null data, but its normal traffic was sometimes labeled Normal and sometimes labeled Benign. Hence, to clean the dataset, we replace the Benign values with the Normal ones. Then, we specified the next step of the attacks in the dataset records. At this point, the next step of each data record is defined as follows:
\begin{itemize}
\item If the current data label is normal, the next step is the first data that is recorded in terms of time after the current data and its source IP address is the same as the current data.
\item If the current data label is not normal, the next step of the attack is the first attack data recorded in terms of time after the current data.
\item Otherwise, the current data is considered the last traffic flow data, which naturally does not define the next step for this data and will be removed from the dataset.
\end{itemize}
Accordingly, the current step label will be replaced by the next step label of the data. The data distribution diagram of different attack classes is demonstrated in Fig. 6. 
\begin{figure*}
\begin{center}
\includegraphics[scale=0.4]{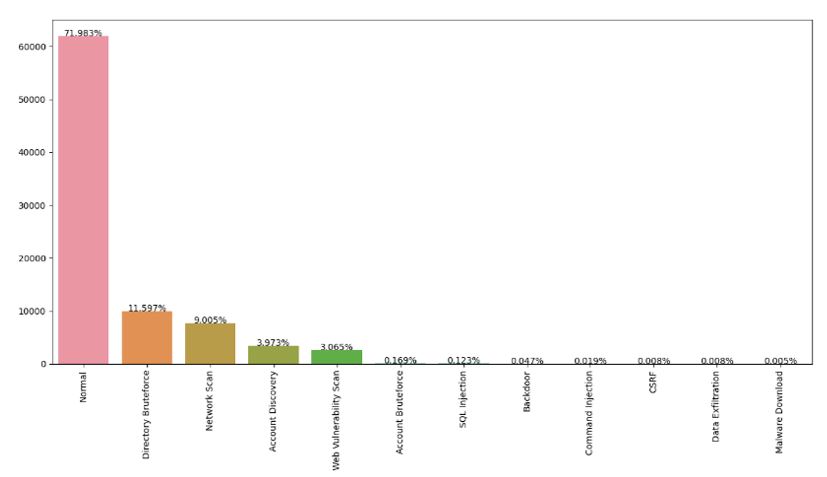}
\small \caption{Distribution of DAPT2020 labels after specification of the next step label}
\label{Figure 6} 
\end{center}
\end{figure*}
\begin{figure*}
\begin{center}
\includegraphics[scale=0.35]{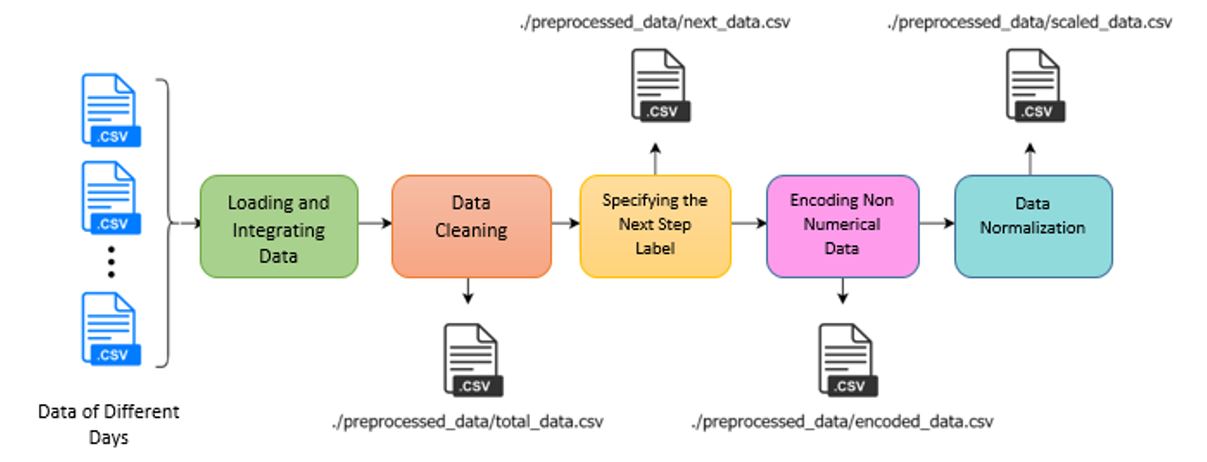}
\small \caption{Dataset Preprocessing Stages}
\label{Figure 7} 
\end{center}
\end{figure*}
\begin{figure*}
\begin{center}
\includegraphics[scale=0.35]{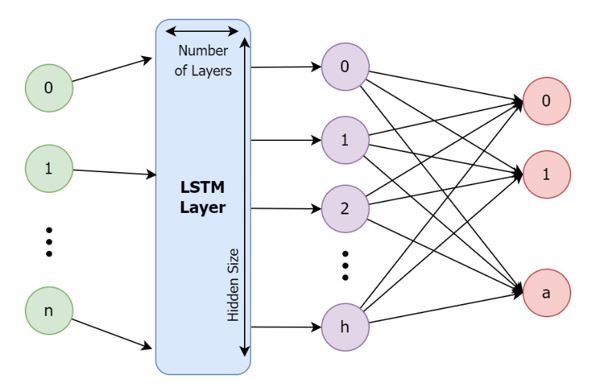}
\small \caption{LSTM Neural Network Scheme}
\label{Figure 8} 
\end{center}
\end{figure*}
\begin{figure*}
\begin{center}
\includegraphics[scale=0.35]{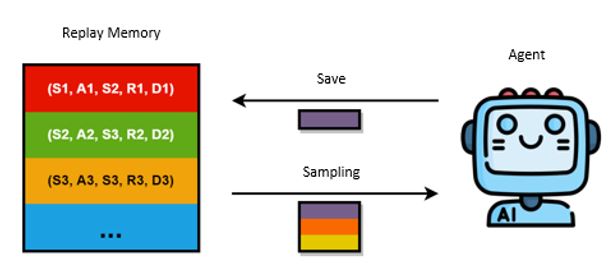}
\small \caption{Interaction of the agent with Replay Memory}
\label{Figure 9} 
\end{center}
\end{figure*}
\begin{figure*}
\begin{center}
\includegraphics[scale=0.35]{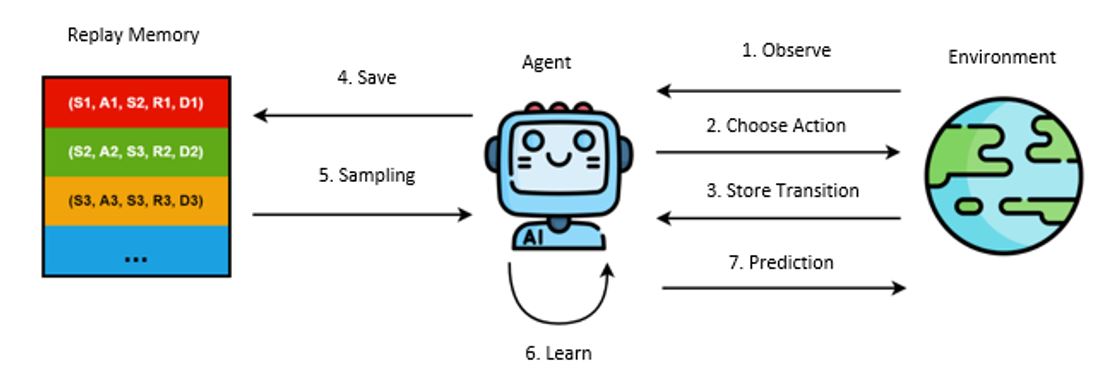}
\small \caption{Interaction of the agent with the environment}
\label{Figure 10} 
\end{center}
\end{figure*}
After specification of the next step label, we encoded non-numerical data as follows:
\begin{itemize}
\item  Source and destination IP address: The IP address is parsed into four separate numbers, each of which naturally ranges from zero to 255.
\item Time label: the time label is stored as "Year-Month-Day Hour : Minute : Second". The time label was coded in Unix time format, which displays the time label as the seconds elapsed from the desired time source, often considered January 1, 1970. 
\item Activity label: This feature is categorical and we encoded them by the One-Hot Encoding method.
\end{itemize}
In the final step of data preprocessing, the data was normalized using formula (2):
$$ z  \ = \ \frac{(x \ - \mu)}{\sigma}       \    \    \    \    \                 (2) $$
In the above formula,$\mu$ is the mean value and $\sigma$ is the standard deviation of data. After this normalization, the mean of the data will be zero and their standard deviation will be one. The stages of our dataset preprocessing are illustrated in Fig. 7. \\
We implemented our proposed approach in Python and using the Pytorch library. We used Mean Squared Error (MSE) for the loss function of the neural network, and Adam optimizer for optimization of the network. \\
For the neural network, the input vector is first given to the LSTM network and the result will be a hidden size vector, as illustrated in Fig. 8. This output is then normalized using Softmax activation function, and then is given to a fully connected layer input network. The number of output nodes in this network is equal to the number of attack steps.\\
Based on Fig. 9, every agent requires memory to remember his observations for learning. \\
The records stored in the memory are comprised of the current state of the agent, the action performed by it, the next state after performing the action, the received reward, and the Done Flag (Done Flag is true if the current data is the last data of training set). \\
To learn from memory, sampling is performed in the batch size. Because of the importance of data time order, samples will be arranged in an ascending order based on the time label and are eventually returned as the output of sampling.\\
The agent employs a neural network, replay memory, and observation of the environment to take actions on the   environment. Then, the agent receives feedback from the environment and remembers it. In the next step, it will sample the memory in the batch size, and learn according to the received samples. \\
Based on Fig. 10, the agent receives observations from the environment and uses $\epsilon - greedy$ policy to select the action. First, it generates a random number in the range [0, 1) and if this number is less than or equal to the $\epsilon$ value, the agent performs a random action. If this number exceeds the $epsilon$ value, it converts the observation factor into a suitable form and gives it as input to the main neural network. The neural network returns the Q-value of each action and selects the action with the highest Q-value. Then, it performs the best action on the environment. 
Afterward, the agent converts the received observations (rewards) from the environment into a suitable format and stores them in his memory. \\
In stage 6 of Fig. 10, the agent learns from the experiences that are stored in the replay memory. It predicts the next action based on the experiences. The difference between predicting and choosing the action is in the use of $\epsilon - greedy$ policy. In prediction, $\epsilon - greedy$ policy is not used, and the action with the largest Q-value is selected. \\
As illustrated in Fig. 5, the learning stage is performed by the main network. If the number of rounds needed to update the target network goes through, all parameters of the main network will be copied to the target network. Finally, the $q_{ref}$ value is calculated as formula (3): 
$$ q_{ref} \ = \ r_t \ + \ \lambda \ \times \ q_{t+1} \ \ \ \ \   (3) $$
Where $q_{t+1}$ is the output of the target network, $\lambda$ is the discount factor and $r_t$ is the rewards of the actions performed.\\
Then, the values $q_t$ and $q_{ref}$ are given as inputs to the main neural network loss function to calculate the loss value. Finally, the loss function will be calculated as formula (4):
         $$  Loss \ = \  \frac{1}{n} \ \Sigma_n (q_t \ - \ q_{ref})^2     \ \ \ \ \      (4) $$
In the next step, the main neural network optimizer performs optimization operations based on the loss function.
In the last step, the $\epsilon$ value will be updated as formula (5):
$$
\epsilon = \left\{ \begin{array}{rcl}
\epsilon - \epsilon_{decrement} & \mbox{if}
& \epsilon > \epsilon_{end} \\
\epsilon & \mbox{if} & \epsilon \leq \epsilon_{end}
\end{array}\right. \ \ \ \ \ \  (5)
$$
\subsection{Experiment with DAPT2020 dataset}
The environment in which the agent tries to learn its behavior is a supervised environment based on DAPT2020 dataset that has been processed in the preprocessing stage of data. \\
We applied the Stratified K-Fold Cross-Validation method to separate the test and training set, which is the modified version of K-Fold Cross-Validation. The difference between these two methods is in the distribution of classes in each split. The Stratified method ensures that the distribution of different classes in each split is equal to the other split. Because the number of data in some classes is very low, with the classic K-Fold Cross-Validation method, the chances of all classes in all splits will be very low. We consider K = 4, so $75 \% $  of the data will be considered for training and $25 \% $ for testing. \\
In DAPT2020, the distribution of classes is unbalanced. Hence, it is difficult to learn data with a low number. One of the methods to solve this problem is oversampling. In this method, the number of data that is very low increases by multiple sampling. \\
In this environment (DAPT2020), oversampling is performed in two stages: 
\begin{itemize}
\item Step 1 – Random Oversampling:\\
In this method, the data with lower numbers are repeated randomly to reach a minimum number for the next stage of oversampling. In this environment, three classes with lower numbers are affected by this stage. 
\item Step 2 – Oversampling by SMOTE method: \\
SMOTE is one of the most widely used methods of oversampling that prevents direct repetition of data like the previous one, by using K-Nearest Neighbor (KNN) algorithm.
For a simple description of SMOTE method, we show an example in two-dimensional space (Fig. 11). Using the KNN method, the neighbors of each data are calculated and lines from the mentioned data to its neighbors are drawn.
\begin{figure}[H]
\begin{center}
\includegraphics[scale=0.4]{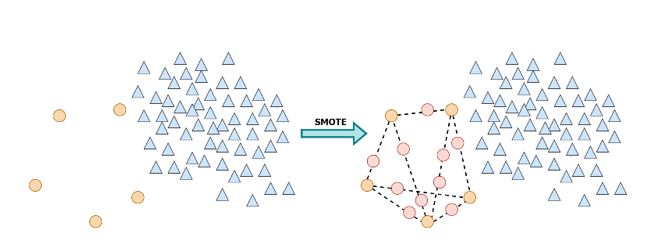}
\small \caption{ SMOTE over-sampling algorithm with K=3}
\label{Figure 11} 
\end{center}
\end{figure}
Finally, the new points on these lines are randomly generated as new data. In this environment, K is considered to be 6 and the seven classes with the lowest number are affected by this stage.
The important point in this section is that oversampling operations are not applied to the whole data and will only be applied to the data of the training set. Therefore, we will not have data that will be present in both the training set and the test set. After applying two stages of oversampling, due to the importance of data time order, they are arranged in an ascending order based on the time label.
\end{itemize}
After oversampling, the distribution of classes is balanced. Hence, the environment (dataset) is ready for agent training. For training the agent, 10 epoch was considered. \\
In each epoch, the agent receives data as an environment observation and selects the next step of the attack in the form of action selection. If the next step of the attack is chosen correctly, it receives a reward +1 and otherwise zero. \\
Then, the agent selects the current state of the environment (current data), the next step of the selected attack (the selected action), the next state of the environment (the next data), the received reward, and the Done Flag as inputs, and stores them in its memory as experience.\\
After completing the training of a split, the test data are given to the agent for prediction and the results are illustrated as Confusion Matrix. The calculated evaluation metrics will be Accuracy, F1, Precision, Recall, and Loss.

\subsection{The effect of Parameters}
Due to the great impact of parameters on the results, the agent was trained with different values and evaluated to assign the best values to them. In the following, we argue about the impact of the parameters. We have performed many experiments but we only state the best results in this section.

\begin{itemize}

\item	Discount Factor:
Unlike the usual reinforcement learning problems that adopt a large discount factor of nearly 1, in this implementation, due to the use of ready datasets instead of the real environment, small amounts of the discount factor, close to zero, make better results, as demonstrated in Fig. 12.
\begin{figure}[H]
\begin{center}
\includegraphics[scale=0.5]{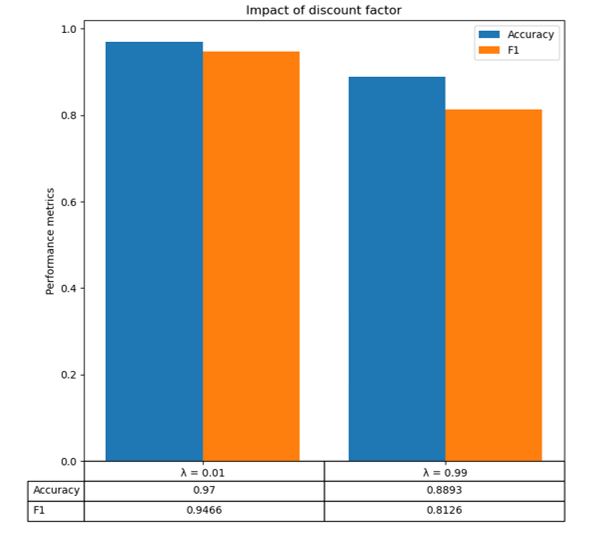}
\small \caption{  The impact of the discount factor on the results }
\label{Figure 12} 
\end{center}
\end{figure}
\item  Final Epsilon: 
The epsilon parameter introduces randomness into the algorithm, forcing us to try different actions. This helps not get stuck in a local optimum. According to the following results (Table 1), if the agent selects randomly $10 \%$ of actions, new and valuable experiences are created for the agent.
\begin{table}[H]
\begin{center}
\includegraphics[scale=0.5]{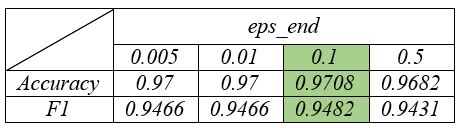}
\small \caption{The impact of final epsilon on the results}
\label{Table 1} 
\end{center}
\end{table}
\item  Maximum memory size: 
Considering that the training data is approximately 78,000 and the following results (Table 2) are obtained, it can be said that if the maximum memory size for the agent is at least equal to the number of training data and the maximum size is twice that, appropriate results will be achieved.  If the memory is less than this amount, it will lose previous experiences. If it exceeds this value, excessively old experiences are also used in the training step. In both cases, they reduce the accuracy of the agent.
\begin{table}[H]
\begin{center}
\includegraphics[scale=0.5]{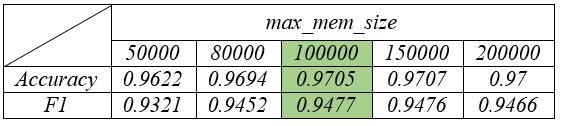}
\small \caption{The impact of final maximum memory size on the results}
\label{Table 2} 
\end{center}
\end{table}
\item  Learning rate: 
Reinforcement learning agents usually learn with a low learning rate, which according to the following results (Table 3), this issue is also established for the desired agent. The appropriate amount for the learning rate is marked in green.
\begin{table}[H]
\begin{center}
\includegraphics[scale=0.5]{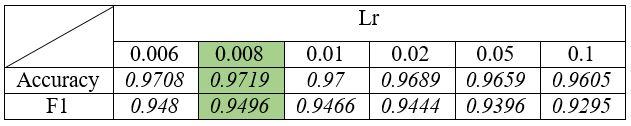}
\small \caption{The impact of the learning rate parameter on the results}
\label{Table 3} 
\end{center}
\end{table}
\item  The number of rounds needed to update the target network: 
Based on the results illustrated in Table 4, in an environment where the agent is training, delay in updating the target network has no impact on the accuracy and performance of the agent training.
\begin{table}[H]
\begin{center}
\includegraphics[scale=0.5]{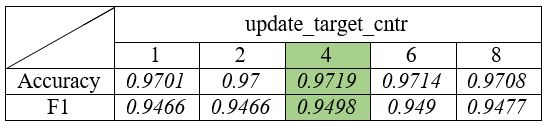}
\small \caption{The impact number of rounds needed to update target network on the results}
\label{Table 4} 
\end{center}
\end{table}
\item  Hidden size: 
As illustrated in Table 5, the more the hidden size increases, the more complexity of the model and the accuracy of the model improves, but it is important to note that with increasing the complexity of the model, the training time also increases significantly, and the risk of over-fitting the model increases. As a result, these parameters are chosen to be suitable both in time and accuracy.
\begin{table}[H]
\begin{center}
\includegraphics[scale=0.5]{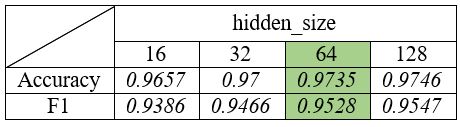}
\small \caption{The impact of hidden size on the results}
\label{Table 5} 
\end{center}
\end{table}
\item  Batch size: 
Based on Table 6, the more increase in the size of the batch, the accuracy of the model increases. Each time more data is given as input to the network, the training time also increases. As a result, the batch size is chosen that is desirable both in terms of time and accuracy. 
\begin{table}[H]
\begin{center}
\includegraphics[scale=0.5]{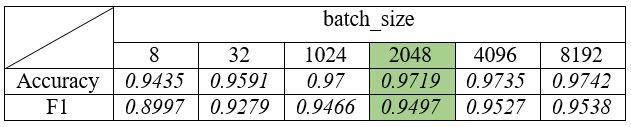}
\small \caption{The impact of batch size on the results}
\label{Table 6} 
\end{center}
\end{table}
Based on the conditions mentioned in the previous sections about training, and the impacts of parameters on the results, we state the results in this section. The values assigned to the parameters are demonstrated in Table 7. 
\begin{table}[H]
\begin{center}
\includegraphics[scale=0.5]{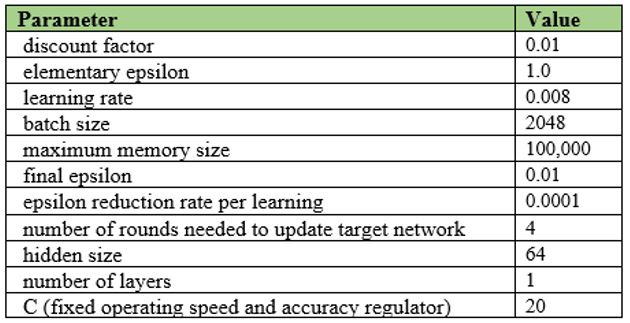}
\small \caption{our proposed system parameters and values}
\label{Table 7} 
\end{center}
\end{table}
As mentioned before, the dataset was divided into four different splits using the Stratified Cross Validation method. To present the results, the mean of split results has been calculated and illustrated in Table 8.
\begin{table}[H]
\begin{center}
\includegraphics[scale=0.5]{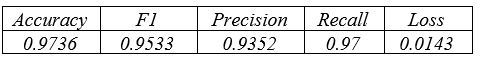}
\small \caption{Evaluation Results}
\label{Table 8} 
\end{center}
\end{table}
For example, the results of a split with details, in the form of a confusion matrix are demonstrated in Fig. 13. 
\begin{figure*}
\begin{center}
\includegraphics[scale=0.5]{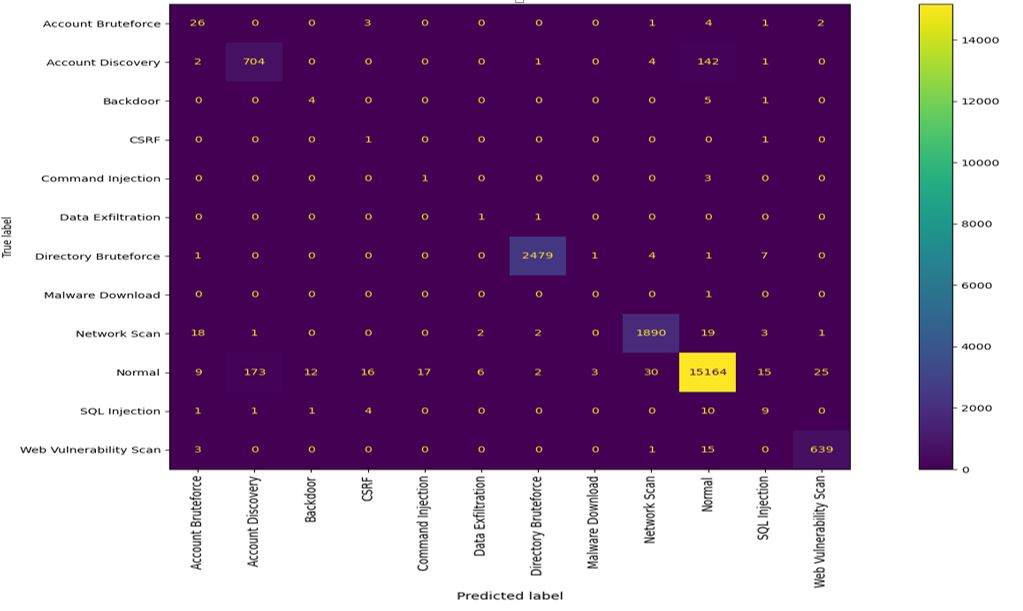}
\small \caption{The confusion matrix related to the sample split}
\label{Figure 13} 
\end{center}
\end{figure*}
\item Time and memory consumption:
Evaluation of the proposed system in terms of time and memory is needed to check the functionality of this system in practical use. The number of data records is visible in table 9. In addition, Memory occupied by data can be seen at different stages in table 10. 
\begin{table}[H]
\begin{center}
\includegraphics[scale=0.5]{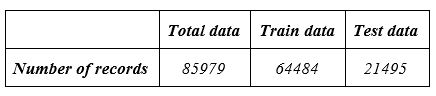}
\small \caption{The Number of Data Records}
\label{Table 9} 
\end{center}
\end{table}
As oversampling is applied only to training data, the blank field is demonstrated in the last row of table 10.
\begin{table}[H]
\begin{center}
\includegraphics[scale=0.4]{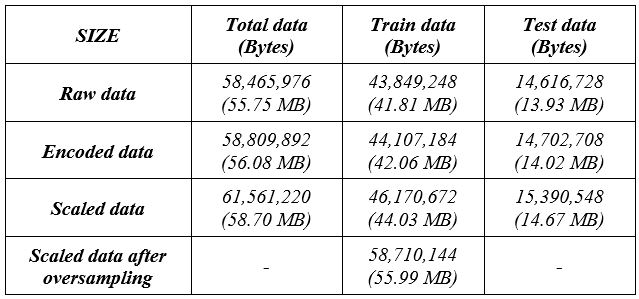}
\small \caption{Memory Consumption}
\label{Table 10} 
\end{center}
\end{table}
The elapsed time for data preprocessing is illustrated in table 11. Moreover, the spent time for both training and test phases can be seen in table 12. 
\begin{table}[H]
\begin{center}
\includegraphics[scale=0.5]{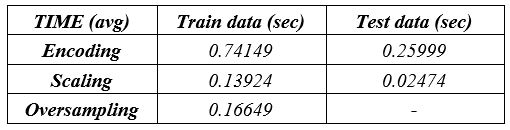}
\small \caption{Elapsed Time}
\label{Table 11} 
\end{center}
\end{table}
\begin{table}[H]
\begin{center}
\includegraphics[scale=0.45]{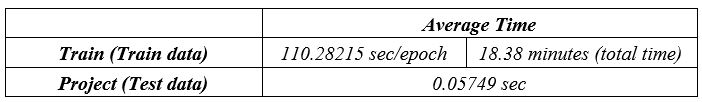}
\small \caption{Spent Time for training and test phases}
\label{Table 12} 
\end{center}
\end{table}
\end{itemize}

We evaluated our proposed deep reinforcement learning system on the DAPT2020 dataset. Based on the evaluations performed on the mentioned dataset, six criteria F1, Accuracy, Precision, Recall, Loss, and average time were obtained, which are 0.9533, 0.9736, 0.9352, 0.97, 0.0143, and 0.05749(seconds) respectively. Although there is no previous research on using reinforcement learning for APT projection, our results compared to the previous supervised and unsupervised methods proposed for multi-step attack projections indicate appropriate functioning. In table 13, we compare the evaluation results of our proposed model and the previous researches. \\ 

\begin{table*}
\begin{center}
\includegraphics[scale=0.7]{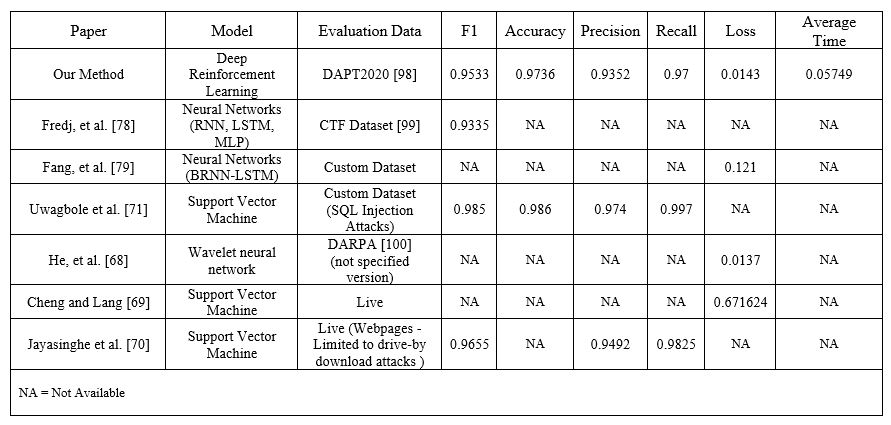}
\small \caption{A comparison of evaluations scores for our method with the previous projection methods}
\label{Table 12} 
\end{center}
\end{table*}

It should be noted that CPU was used for all software resources, and GPU has not been used. In addition, the employed processor model was Intel Core i7-4790K $@$ 4.00GHz.
\pagebreak
\section{Conclusion}
The highest level in Endsley's situation awareness model is called projection, when the status of elements in the environment in the near future is predicted. \\
In this paper, we presented a deep reinforcement learning system to project the next step of APTs. As there exists some relation between attack steps, we employed LSTM to approximate the best action of each state. In our proposed system, based on the current situation, we project the next steps of APT threats. This contribution helps to select appropriate and timely countermeasures against APTs.\\
In our proposed system, the agent learns based on the dataset of APTs called DAPT2020 and projects the next step of the attacks. Since the DAPT2020 has about $96 \% $ normal traffic and $4 \% $  attack traffic, oversampling was performed on the DAPT2020 to increase the amount of attack traffic. To train and test the system, the dataset was divided into 4 splits based on the Stratified Cross-Validation method. Then we have considered three splits for training and one split for testing and evaluation. The agent was trained and tested based on the mentioned descriptions. \\
We have evaluated our proposed deep reinforcement learning system on the DAPT2020 dataset. Based on the evaluations performed on the mentioned dataset, six criteria F1, Accuracy, Precision, Recall, Loss, and average time were obtained, which are 0.9533, 0.9736, 0.9352, 0.97, 0.0143, and 0.05749(seconds) respectively. Although there is no previous research on using reinforcement learning for APT projection, our results compared to the previous supervised and unsupervised methods proposed for multi-step attack projections indicate appropriate functioning. \\ 
As stated in the Experiments section, the average time for predicting the next step of APTs is 0.05749 seconds. We can employ the proposed system in real-time intrusion detection and prevention systems. For future work, it would be worthwhile to employ our proposed method in a real-time intrusion detection and prevention system. \\
To reduce the time of learning, testing, and evaluation, we employed APT datasets instead of interacting with the environment. Although employing datasets increases the speed of learning and testing, interaction with the environment is suitable for predicting unknown APTs. In our proposed model the agent can employ datasets or interact with the environment to project the next step of APTs. For future work, it would be worthwhile to interact with the real environment to test for APT projection and identifying unknown threats.\\
Moreover, due to the great impact of parameters on the results, the agent was trained with different values and it was also evaluated for the assignment of the best values to them. Employing evolutionary algorithms such as genetic algorithms to determine the best values for our proposed system parameters would be worthwhile too.\\
Another future work is the automatic determination of countermeasures based on the predicted steps of attacks and other important information in cybersecurity situation awareness. For example, the projected attack steps, their severity, and the importance of the target assets are effective components in the automatic determination of countermeasures, which we intend to propose in our future research.
\section{Bibliography}
\small

\noindent [1].  MR. Endsley, \textit{Toward a Theory of Situation Awareness in Dynamic Systems}, Human factors, vol. 37, no.1, 1995, pp. 32-64. \\

\noindent [2].  R. S. Sutton, A. G. Barto, \textit{Introduction to reinforcement learning}, Cambridge: MIT press, vol. 135, 1998.  \\

\noindent [3]. F. Cohen, \textit{Information System Defences: a Preliminary Classification Scheme}, Computers $\&$ Security, 1997 Jan 1, vol. 16, no. 2, pp. 94-114.\\
 
\noindent [4]. F. Cohen, \textit{Simulating Cyber Attacks, Defences, and Consequences}, Computers $\&$ Security, 1999 Jan 1, vol. 18, no. 6, pp. 479-518. \\

\noindent [5]. JD. Howard, TA. Longstaff, \textit{A Common Language for Computer Security Incidents}, Sandia National Lab.(SNL-NM), Albuquerque, NM (United States), Sandia National Lab.(SNL-CA), Livermore, CA (United States), 1998 Oct 1.\\

\noindent [6]. H. Debar, M. Dacier, A. Wespi, \textit{Towards a Taxonomy of Intrusion-Detection Systems}, Computer networks, 1999 Apr 23, vol. 31, no. 8, pp. 805-822.\\

\noindent [7]. A. Chakrabarti, G. Manimaran, \textit{Internet Infrastructure Security: A Taxonomy}, IEEE network, 2002 Dec 16, vol. 16, no. 6, pp. 13-21.\\

\noindent [8]. P. Ning, Y. Cui, DS. Reeves, \textit{Analyzing Intensive Intrusion Alerts via Correlation}, In International Workshop on Recent Advances in Intrusion Detection 2002 Oct 16, pp. 74-94, Springer, Berlin, Heidelberg.\\

\noindent [9]. P. Ning, D. Xu, CG. Healey, RS. Amant, \textit{Building Attack Scenarios through Integration of Complementary Alert Correlation Method}, In NDSS 2004 Feb 4, vol. 4, pp. 97-111.\\
 
\noindent [10]. S. Cheung, U. Lindqvist, MW. Fong, \textit{Modeling Multistep Cyber Attacks for Scenario Recognition}, In Proceedings DARPA Information Survivability Conference and Exposition 2003 Apr 22, vol. 1, pp. 284-292.\\
 
\noindent [11]. F. Valeur, et al., \textit{Comprehensive Approach to Intrusion Detection Alert Correlation}, IEEE Transactions on dependable and secure computing, 2004 Dec 13, vol. 1, no. 3, pp. 146-69.\\

\noindent [12]. S. Noel, E. Robertson, S. Jajodia, \textit{Correlating Intrusion Events and Building Attack Scenarios Through Attack Graph Distances}, In $20^{th}$ Annual Computer Security Applications Conference 2004 Dec 6 pp. 350-359. \\

\noindent [13]. ST. King, et al., \textit{Enriching Intrusion Alerts Through Multi-Host Causality}, In NDSS 2005 Feb. \\

\noindent [14]. L. Wang, A. Liu, S. Jajodia, \textit{Using Attack Graphs for Correlating, Hypothesizing, and Predicting Intrusion Alerts}, Computer Communications, 2006 Sep 5, vol. 29, no. 15, pp. 2917-2933. \\

\noindent [15]. A. Stotz, M. Sudit, \textit{INformation Fusion Engine for Real-Time Decision-Making (INFERD): A Perceptual System for Cyber Attack Tracking}, In 2007 $10^{th}$ International Conference on Information Fusion 2007 Jul 9, pp. 1-8, IEEE.\\

\noindent [16]. SJ. Yang, et al., \textit{High Level Information Fusion for Tracking and Projection of Multistage Cyber Attacks}, Information Fusion, 2009 Jan 1, vol. 10, no. 1, pp. 107-21.\\

\noindent [17]. M. Husák, J. Kašpar, \textit{AIDA Framework: Real-time Correlation and Prediction of Intrusion Detection Alerts}, In Proceedings of the $14^{th}$ international conference on availability, reliability and security 2019 Aug 26, pp. 1-8.\\

\noindent [18]. X. Qin, W. Lee, \textit{Attack Plan Recognition and Prediction Using Causal Networks}, In $20^{th}$ Annual Computer Security Applications Conference 2004 Dec 6, pp. 370-379. \\

\noindent [19]. S. Noel, S. Jajodia, \textit{Advanced Vulnerability Analysis and Intrusion Detection Through Predictive Attack Graphs}, Critical Issues in C4I, Armed Forces Communications and Electronics Association (AFCEA) Solutions Series, International Journal of Command and Control, 2009 May. \\

\noindent [20]. C. Phillips, LP. Swiler, \textit{A Graph-based System for Network-Vulnerability Analysis}, In Proceedings of the 1998 workshop on New security paradigms 1998 Jan 1, pp. 71-79. \\

\noindent [21]. T. Tidwell, et al., \textit{Modeling Internet Attacks}, In Proceedings of the 2001 IEEE Workshop on Information Assurance and Security, vol. 59, 2001. \\

\noindent [22]. K. Daley, R. Larson, J. Dawkins, \textit{A Structural Framework for Modeling Multi-Stage Network Attacks}, In Proceedings of International Conference on Parallel Processing, pp. 5–10, 2002. \\

\noindent [23]. S. Vidalis, A. Jones, \textit{Using Vulnerability Trees for Decision Making in Threat Assessment}, Technical Report CS-03-2, University of Glamorgan, School of Computing, June 2003. \\

\noindent [24]. T. Hughes, O. Sheyner, \textit{Attack Scenario Graphs for Computer Network Threat Analysis and Prediction}, Complexity, vol. 9, no. 2, pp. 15–18, 2003.\\

\noindent [25] C. J. Chung, et al., \textit{NICE: Network Intrusion Detection and Countermeasure Selection in Virtual Network Systems}, IEEE Transactions on Dependable and Secure Computing, vol. 10, no. 4, pp. 198–211, July 2013. \\

\noindent [26] I. Kotenko and A. Chechulin, \textit{A Cyber Attack Modeling and Impact Assessment Framework}, in 2013 $5^{th}$ International Conference on Cyber Conflict (CYCON 2013), June 2013, pp. 1–24. \\

\noindent [27] P. Cao, et al. \textit{Preemptive Intrusion Detection}, in Proceedings of the 2014 Symposium and Bootcamp on the Science of Security, New York, NY, USA: ACM, 2014, pp. 1-21. \\

\noindent [28] P. Cao, et al., \textit{Preemptive Intrusion Detection: Theoretical Framework and Real-world Measurements}, in Proceedings of the 2015 Symposium and Bootcamp on the Science of Security, New York, NY, USA: ACM, 2015, pp. 1-5. \\

\noindent [29] A. A. Ramaki, M. Amini, and R. E. Atani, \textit{RTECA: Real Time Episode Correlation Algorithm for Multi-step Attack Scenarios Detection}, Computers $\&$ Security, vol. 49, no. Supplement C, pp. 206 – 219, 2015. \\

\noindent [30] M. GhasemiGol, A. Ghaemi-Bafghi, and H. Takabi, \textit{A Comprehensive Approach for Network Attack Forecasting}, Computers $\&$ Security, vol. 58, pp. 83 – 105, 2016. \\

\noindent [31] M. GhasemiGol, H. Takabi, and A. Ghaemi-Bafghi, \textit{Foresight Model for Intrusion Response Management}, Computers $\&$ Security, vol. 62, pp. 73 – 94, 2016. \\

\noindent [32] N. Polatidis, et al., \textit{Recommender Systems Meeting Security: From Product Recommendation to Cyber-Attack Prediction}, in Engineering Applications of Neural Networks, Cham: Springer International Publishing, 2017, pp. 508– 519. \\

\noindent [33] N. Polatidis, et al., \textit{From Product Recommendation to Cyber-Attack Prediction: Generating Attack Graphs and Predicting Future Attacks}, Evolving Systems, May 2018.\\

\noindent [34] X. Qin and W. Lee, \textit{Attack Plan Recognition and Prediction Using Causal Networks}, in Computer Security Applications Conference, $20^{th}$ Annual, Dec 2004, pp. 370–379. \\

\noindent [35] J. Wu, L. Yin, and Y. Guo, \textit{Cyber Attacks Prediction Model Based on Bayesian Network}, in Parallel and Distributed Systems (ICPADS), 2012 IEEE $18^{th}$ International Conference on, Dec 2012, pp. 730–731. \\

\noindent [36] A. A. Ramaki, M. Khosravi-Farmad, and A. G. Bafghi, \textit{Real time alert correlation and prediction using Bayesian networks}, in Information Security and Cryptology (ISCISC), 2015 $12^{th}$ International Iranian Society of Cryptology Conference on. IEEE, 2015, pp. 98–103. \\

\noindent [37] A. Okutan, S. J. Yang, and K. McConky, \textit{Predicting Cyber Attacks with Bayesian Networks Using Unconventional Signals}, in Proceedings of the $12^{th}$ Annual Conference on Cyber and Information Security
Research, ACM, 2017, pp. 1-13. \\

\noindent [38] K. Huang, et al., \textit{Assessing the Physical Impact of Cyberattacks on Industrial Cyber-Physical Systems}, IEEE Transactions on Industrial Electronics, vol. 65, no. 10, pp. 8153 – 8162, Oct 2018. \\

\noindent [39]. D. S. Fava, S. R. Byers, S. J. Yang, \textit{Projecting Cyberattacks Through Variable-Length Markov Models}, IEEE Transactions on Information Forensics and Security, vol. 3, no. 3, pp. 359–369, September 2008. \\

\noindent [40]. P. Jacquet, W. Szpankowski, I. Apostol, \textit{A Universal Predictor Based on Pattern Matching} , IEEE Transactions on Information Theory, vol. 48, no. 6, pp. 1462–1472, June 2002. \\

\noindent [41]. C. R. Shalizi, K. L. Shalizi, \textit{Blind Construction of Optimal Nonlinear Recursive Predictors for Discrete Sequences}, In Proceedings of the $20^{th}$ Conference on Uncertainty in Artificial Intelligence, pp. 504–511, 2004. \\

\noindent [42] H. Farhadi, M. AmirHaeri, and M. Khansari, \textit{Alert Correlation and Prediction Using Data Mining and HMM}, ISeCure, vol. 3, no. 2, 2011. \\

\noindent [43] A. S. Sendi, M. Dagenais, and M. Jabbarifar, \textit{Real Time Intrusion Prediction based on Optimized Alerts with Hidden Markov Model}, Journal of Networks, vol. 7, no. 2, 2012. \\

\noindent [44] S. Shin, et al., \textit{Advanced Probabilistic Approach for Network Intrusion Forecasting and Detection}, Expert Systems with Applications, vol. 40, no. 1, pp. 315 – 322, 2013.\\

\noindent [45] Y. Zhang, D. Zhao, and J. Liu, \textit{The Application of Baum-Welch Algorithm in Multistep Attack}, The Scientific World Journal, vol. 2014, 2014. \\

\noindent [46] H. A. Kholidy, A. Erradi, and S. Abdelwahed, \textit{Attack Prediction Models for Cloud Intrusion Detection Systems}, in Artificial Intelligence, Modelling and Simulation (AIMS), 2014 2nd International Conference on, Nov 2014, pp. 270–275. \\

\noindent [47] H. A. Kholidy, et al., \textit{A Finite State Hidden Markov Model for Predicting Multistage Attacks in Cloud Systems}, in Dependable, Autonomic and Secure Computing (DASC), 2014 IEEE $12^{th}$ International Conference on, Aug 2014, pp. 14–19. \\

\noindent [48] H. A. Kholidy, et al., \textit{A Finite Context Intrusion Prediction Model for Cloud Systems with a Probabilistic Suffix Tree}, in Modelling Symposium (EMS), 2014 European, Oct 2014, pp. 526–531. \\

\noindent [49] S. Abraham and S. Nair, \textit{Exploitability Analysis using Predictive Cybersecurity Framework}, in 2015 IEEE 2nd International Conference on Cybernetics (CYBCONF), June 2015, pp. 317–323. \\

\noindent [50] A. Bar, et al., \textit{Identifying Attack Propagation Patterns in Honeypots Using Markov Chains Modeling and Complex Networks Analysis}, in Software Science, Technology and Engineering (SWSTE), 2016 IEEE International Conference on. IEEE, 2016, pp. 28–36. \\

\noindent [51] B. Ariel, et al., \textit{Scalable Attack Propagation Model and Algorithms for Honeypot Systems}, In 2016 IEEE International Conference on Big Data (Big Data), pp. 1130-1135. IEEE, 2016. \\

\noindent [52]. T. Chadza, K. G. Kyriakopoulos, S. Lambotharan, \textit{Analysis of Hidden Markov Model Learning Algorithms for the Detection and Prediction of Multi-Stage Network Attacks}, Future Generation Computer Systems 108, 2020, pp. 636-649. \\

\noindent [53]. Z. Dongmei, H. Wang, S. Geng. \textit{Compound Attack Prediction Method Based on Improved Algorithm of Hidden Markov Model}, Journal of Web Engineering, 2020, pp. 244-263. \\

\noindent [54]. I. Sanjana, M. Paraye, D. Ambawade, \textit{Enhancing Multi-Step Attack Prediction using Hidden Markov Model and Naive Bayes}, In 2020 International Conference on Electronics and Sustainable Communication Systems (ICESC), pp. 36-44. IEEE, 2020. \\

\noindent [55] M. Abdlhamed, et al., \textit{A System for Intrusion Prediction in Cloud Computing}, in Proceedings of the International Conference on Internet of Things and Cloud Computing, New York, NY, USA: ACM, 2016, pp. 1-35. \\

\noindent [56] V. Lisy, et al., \textit{Gametheoretic Approach to Adversarial Plan Recognition}, in ECAI, 2012, pp. 546–551. \\

\noindent [57] R. Pibil, et al., \textit{Game Theoretic Model of Strategic Honeypot Selection in Computer Networks}, in Decision and Game Theory for Security, Springer, 2012, pp. 201–220.\\

\noindent [58]. W. Lee, S. J. Stolfo, P. K. Chan, \textit{Learning Patterns from Unix Process Execution Traces for Intrusion Detection}, In Proceedings of the workshop on AI Approaches to Fraud Detection and Risk Management, pp. 50–56, 1997.\\

\noindent [59]. T. Lane, C. Brodley, \textit{Temporal Sequence Learning and Data Reduction for Anomaly Detection}, ACM Transactions on Information and System Security, vol. 2, pp. 295–331, 1999. \\

\noindent [60]. N. Ye, Y. Zhang, C. M. Borror, \textit{Robustness of the Markov-Chain Model for Cyber-Attack Detection}, IEEE Transactions on Reliability, vol. 53, pp. 116–123, 2004. \\

\noindent [61] P. Shao, et al., \textit{ J. Lu, R. K. Wong, and W. Yang, A Transparent Learning Approach for Attack Prediction based on User Behavior Analysis}, in Information and Communications Security, Cham: Springer International Publishing, 2016, pp. 159–172. \\

\noindent [62] Y. Liu, et al., \textit{Cloudy with a Chance of Breach: Forecasting Cyber Security Incidents}, in USENIX Security Symposium, 2015, pp. 1009–1024.\\

\noindent [63] R. Zheng, et al., \textit{A Strategy of Network Security Situation Autonomic Awareness}, in Network Computing and Information Security. Springer, 2012, pp. 632–639.\\

\noindent [64] F. Chen, et al., \textit{The Network Security Situation Predicting Technology based on the Small-World Echo State Network}, in Software Engineering and Service Science (ICSESS), 2013 $4^{th}$ IEEE International Conference on. IEEE, 2013, pp. 377–380.\\

\noindent [65] Y. Zhang, et al., \textit{Network Security Situation Prediction Based on BP and RBF Neural Network}, Berlin, Heidelberg: Springer Berlin Heidelberg, 2013, pp. 659–665.\\

\noindent [66] W. Xing-zhu, \textit{Network Intrusion Prediction Model based on RBF Features Classification}, International Journal of Security and Its Applications, vol. 10, no. 4, pp. 241–248, 2016. \\

\noindent [67] H. Zhang, et al., \textit{A Network Security Situation Prediction Model based on Wavelet Neural Network with Optimized Parameters}, Digital Communications and Networks, vol. 2, no. 3, pp. 139 – 144, 2016, advances in Big Data.\\

\noindent [68] F. He, et al., \textit{Mixed Wavelet- Based Neural Network Model for Cyber Security Situation Prediction Using MODWT and Hurst Exponent Analysis}, in Network and System Security, Cham: Springer International Publishing, 2017, pp. 99–111.\\

\noindent [69] X. Cheng, S. Lang, \textit{Research on Network Security Situation Assessment and Prediction}, in Computational and Information Sciences (ICCIS), 2012 Fourth International Conference on. IEEE, 2012, pp. 864–867.\\

\noindent [70] G. K. Jayasinghe, J. S. Culpepper, and P. Bertok, \textit{Efficient and Effective Realtime Prediction of Drive-by Download Attacks}, Journal of Network and Computer Applications, vol. 38, pp. 135 – 149, 2014.\\

\noindent [71] S. O. Uwagbole, W. J. Buchanan, and L. Fan, \textit{Applied Machine Learning Predictive Analytics to SQL Injection Attack Detection and Prevention}, in 2017 IFIP/IEEE Symposium on Integrated Network and Service Management (IM), May 2017, pp. 1087–1090. \\

\noindent [72] U. S. Ogbomon, W. J. Buchanan, and L. Fan, \textit{An Applied Pattern-Driven Corpus to Predictive Analytics in Mitigating SQL Injection Attack}, In 2017 Seventh International Conference on Emerging Security Technologies (EST), pp. 12-17. IEEE, 2017. \\

\noindent [73] C. Fachkha, et al., \textit{Investigating the Dark Cyberspace: Profiling, Threat-based Analysis and Correlation}, in 2012 $7^{th}$ International Conference on Risks and Security of Internet and Systems (CRiSIS), Oct 2012. \\

\noindent [74] Y.H. Kim, W. H. Park, \textit{A Study on Cyber Threat Prediction based on Intrusion Detection Event for APT Attack Detection}, Multimedia Tools and Applications, vol. 71, no. 2, pp. 685–698, Jul 2014.  \\

\noindent [75] M. Husak, J. Kaspar, \textit{Towards Predicting Cyber Attacks Using Information Exchange and Data Mining}, in Proceedings of 2018 International Wireless Communications and Mobile Computing Conference (IWCMC), 2018, (to appear). \\

\noindent [76] K. Soska, N. Christin, \textit{Automatically Detecting Vulnerable Websites before They Turn Malicious}, in USENIX Security Symposium, 2014, pp. 625–640. \\

\noindent [77] K. Veeramachaneni, et al., \textit{AI2: Training a Big Data Machine to Defend}, in 2016 IEEE 2nd International Conference on Big Data Security on Cloud (BigDataSecurity), IEEE International Conference on High Performance and Smart Computing (HPSC), and IEEE International Conference on Intelligent Data and Security (IDS), April 2016, pp. 49–54.\\

\noindent [78]. B. Fredj, et al., \textit{CyberSecurity Attack Prediction: a Deep Learning Approach}, In $13^{th}$ International Conference on Security of Information and Networks, pp. 1-6. 2020. \\

\noindent [79]. X. Fang, et al., \textit{A Deep Learning Framework for Predicting Cyber Attacks Rates}, EURASIP Journal on Information security 2019, no. 1, 2019, pp. 1-11. \\

\noindent [80] H. Park, et al., \textit{Cyber Weather Forecasting: Forecasting Unknown Internet Worms Using Randomness Analysis}, in Information Security and Privacy Research. Berlin, Heidelberg: Springer Berlin Heidelberg, 2012, pp. 376–387.\\

\noindent [81] Z. Zhan, M. Xu, and S. Xu, \textit{Characterizing Honeypot-Captured Cyber Attacks: Statistical Framework and Case Study}, IEEE Transactions on Information Forensics and Security, vol. 8, no. 11, pp. 1775–1789, Nov 2013. \\

\noindent [82] A. Silva, et al., \textit{PRBS/EWMA based Model for Predicting Burst Attacks (Brute Froce, DoS) in Computer Networks}, in Ninth International Conference on Digital Information Management (ICDIM 2014), Sept 2014, pp. 194–200. \\

\noindent [83] V. Abaeian, et al., \textit{Intrusion Detection Forecasting Using Time Series for Improving Cyber Defence}, International Journal of Intelligent Systems and Applications in Engineering, 2015 Mar 3, vol. 3, no. 1, pp. 28-33. \\

\noindent [84] TR. Pillai, et al., \textit{Predictive Modeling for Intrusions in Communication Systems Using GARMA and ARMA Models}, In 2015 $5^{th}$ National Symposium on Information Technology: Towards New Smart World (NSITNSW) 2015 Feb 17, pp. 1-6.\\

\noindent [85] J. Freudiger, et al., \textit{Controlled Data Sharing for Collaborative Predictive Blacklisting}, In International Conference on Detection of Intrusions and Malware, and Vulnerability Assessment 2015 Jul 9, pp. 327-349. \\

\noindent [86] YZ. Chen, et al., \textit{Spatiotemporal Patterns and Predictability of Cyberattacks}, PloS one. 2015 May 20, vol. 10, no. 5. \\

\noindent [87] Z. Zhan, M. Xu, S. Xu, \textit{Predicting Cyber Attack Rates with Extreme Values}, IEEE Transactions on Information Forensics and Security, 2015 Apr 13, vol. 10, no. 8, pp. 1666-77. \\

\noindent [88] P. Sokol, A. Gajdoš, \textit{Prediction of Attacks against Honeynet based on Time Series Modeling}, In Proceedings of the Computational Methods in Systems and Software 2017 Sep 12, Springer, Cham, pp. 360-371.\\

\noindent [89] D. Werner, S. Yang, K. McConky, \textit{Time Series Forecasting of Cyber Attack Intensity}, In Proceedings of the $12^{th}$ Annual Conference on cyber and information security research 2017 Apr 4, pp. 1-3.\\

\noindent [90] S. Dowling, M. Schukat, H. Melvin, \textit{Using Analysis of Temporal Variances within a Honeypot Dataset to Better Predict Attack Type Probability}, In 2017 $12^{th}$ International Conference for Internet Technology and Secured Transactions (ICITST) 2017 Dec 11, pp. 349-354. \\

\noindent [91] A. Okutan, et al., \textit{POSTER: Cyber Attack Prediction of Threats from Unconventional Resources (CAPTURE)}, In Proceedings of the 2017 ACM SIGSAC Conference on Computer and Communications Security 2017 Oct 30, pp. 2563-2565.\\

\noindent [92] Z. Lin, et al., \textit{The Prediction Algorithm of Network Security Situation based on Grey Correlation Entropy Kalman Filtering}, In 2014 IEEE $7^{th}$ Joint International Information Technology and Artificial Intelligence Conference 2014 Dec 20, pp. 321-324. \\

\noindent [93] YB. Leau, S. Manickam, \textit{A Novel Adaptive Grey Verhulst Model for Network Security Situation Prediction}, International Journal of Advanced Computer Science $\&$ Applications, 2016, vol. 1, no. 7, pp. 90-5.\\

\noindent [94] YB. Leau, S. Manickam, \textit{An Enhanced Adaptive Grey Verhulst Prediction Model for Network Security Situation}, International Journal of Computer Science and Network Security (IJCSNS), 2016 May 1, vol. 16, no. 5.\\

\noindent [95] MK. Daly, \textit{Advanced Persistent Threat}, Usenix, Nov. 2009 Nov 4,4(4):2013-2016. \\

\noindent [96] NC. Luong, et al. \textit{Applications of Deep Reinforcement Learning in Communications and Networking: A Survey},  IEEE Communications Surveys $\&$ Tutorials, vol. 21, no. 4, 2019, pp. 3133-74.\\

\noindent [97] V. François-Lavet, et al., \textit{An introduction to deep reinforcement learning}, Foundations and Trends in Machine Learning, vol. 11, no. 3-4, 2018, pp. 219-354.\\

\noindent [98] S. Myneni, et al. \textit{DAPT2020-Constructing a Benchmark Dataset for Advanced Persistent Threats}, In International Workshop on Deployable Machine Learning for Security Defense, 2020 Aug 24 (pp. 138-163). Springer, Cham.\\

\noindent [99] DEFCON - Hacking Conference Home: https://www.defcon.org/ \\

\noindent [100] Datasets - MIT Lincoln Laboratory: https://www.ll.mit.edu/r-d/datasets 

\end{document}